# Wireless Sensor/Actuator Network Design for Mobile Control Applications

Feng Xia [1,2], Yu-Chu Tian [2], Yanjun Li [1] and Youxian Sun [1]

[1] State Key Laboratory of Industrial Control Technology, Zhejiang University,
  Hangzhou 310027, China
  E-mail: f.xia@ieee.org.
[2] Faculty of Information Technology, Queensland University of Technology,
  GPO Box 2434, Brisbane QLD 4001, Australia
  E-mail: y.tian@qut.edu.au.

**Abstract:** Wireless sensor/actuator networks (WSANs) are emerging as a new generation of sensor networks. Serving as the backbone of control applications, WSANs will enable an unprecedented degree of distributed and mobile control. However, the unreliability of wireless communications and the real-time requirements of control applications raise great challenges for WSAN design. With emphasis on the reliability issue, this paper presents an application-level design methodology for WSANs in mobile control applications. The solution is generic in that it is independent of the underlying platforms, environment, control system models, and controller design. To capture the link quality characteristics in terms of packet loss rate, experiments are conducted on a real WSAN system. From the experimental observations, a simple yet efficient method is proposed to deal with unpredictable packet loss on actuator nodes. Trace-based simulations give promising results, which demonstrate the effectiveness of the proposed approach.

# 1. Introduction

Recent advances in pervasive computing, communication and sensing technologies are leading to the emergence of wireless sensor/actuator networks (WSANs) [1,2]. A WSAN is a distributed system of sensor nodes and actuator nodes that are interconnected over wireless links. Sensors gather information about the physical world, e.g., the environment or physical systems, and transmit the collected data to controllers/actuators through single-hop or multi-hop communications. From the received information, the controllers/actuators perform actions to change the behaviour of the environment or physical systems. In this way, remote, distributed interactions with the physical world are facilitated. Depending on the type of the target application, nodes in a WSAN can be either stationary or mobile. In many situations, however, sensor nodes are stationary whereas actuator nodes, e.g., mobile robots and unmanned aerial vehicles, are mobile. Sensor nodes are usually low-cost, low-power, small devices equipped with limited sensing, data processing and wireless communication capabilities, while actuator nodes typically have stronger computation and communication powers and more energy budget that allows longer battery life [3]. Regardless, resource constraints apply to both sensors and actuators.

WSANs are not just an enhancement or complement to the intensively-investigated wireless sensor networks (WSNs) [4-8], but go beyond. They are a new generation of sensor networks [2,3]. While WSANs and WSNs share many common considerations concerning network design, such as reliability, connectivity, scalability and energy efficiency, the coexistence of sensors and actuators in WSANs causes substantial difference between these two types of networks. Applications in which some actions are introduced for the purpose of enhancing the monitoring capability of the sensor networks do not embody the essential characteristics of WSANs. On the contrary, actuators in a WSAN should be an integral part of the network and perform actions interacting with the physical world. As a consequence, WSANs have the ability to change the physical world, but WSNs do not. In WSNs, power consumption is generally the primary concern; however, this may not be the case in some WSANs where meeting the real-time, reliable communication requirements may be more important [9].

Although there are many situations in which only WSNs are required, for example, environment monitoring, product quality monitoring, and the like, there are an increasing number of applications that necessitate the use of actuators along with sensors [10-12]. That is, the network system needs to interact with the physical system or environment. Examples of application areas of WSANs include disaster relief operations, intelligent building, home automation, smart spaces, pervasive computing systems, and cyber-physical systems.

Because of the use of both sensors and actuators, WSANs, by definition, exploit the methodology of feedback, which has been recognized as the central element of control systems [13]. The advent of WSANs has the potential to revolutionarily promote existing control applications. It can be envisioned that WSANs will become the backbone of many control applications enabling an unprecedented degree of distributed control. The use of WSANs in control applications has many advantages compared to wired solutions, which are dominant at the moment [14,15]. For instance, WSANs allow more flexible installation and maintenance, fully mobile operation, and monitoring and control of equipments in hazardous and previously difficult-to-access environments. Another important factor that instigates the deployment of WSANs is their relatively low costs [11].

Despite many advantages, WSANs also raise challenges for control applications. Wireless channels have adverse properties, such as path loss, multi-path fading, adjacent channel interference, Doppler shifts, and half-duplex operations [16]. WSANs are known to be notoriously unpredictable and inherently unreliable. This is especially true in the case of low-power communications and in the presence of node mobility. With these characteristics, the quality of service (QoS) of the network cannot be always guaranteed. A natural result is that control applications will suffer from time-varying delay and packet loss, both of which could significantly degrade the control performance, or even cause system instability. Therefore, WSANs must be well designed when deployed to control applications.

The design of WSANs featuring node mobility is investigated in this paper for control applications. The overall goal is to enhance the reliability of WSANs so that the required performance of control applications is guaranteed in dynamic, lossy environments. In particular, our focus is on dealing with unpredictable packet loss caused by unreliable link quality in mobile WSANs, without considering the effects of time-varying delay. The reasons behind this choice of focus can be briefly explained as follows. Firstly, since a packet loss can be equivalently regarded as a delay with a magnitude of infinity, from the viewpoint of control, packet loss is a factor that often has more significant impact on the resulting control performance than delay. Secondly, several methods have been presented in the literature to cope with time-varying delay in control loops closed over wireless sensor networks [17-21], while the packet loss problem that arises in WSANs is yet to be investigated.

An application-level design methodology will be developed for WSANs based on an experimental study of the link quality properties and a compensation method for packet loss. The main contributions of this paper include:

- The link quality of WSANs is characterized in terms of packet loss rate through experiments on a real deployment. The experimental results provide important insight into how the WSAN should be designed from the application point of view. Moreover, they are also of great value to the design and evaluation of sensor network protocols and algorithms.
- A simple yet efficient method is developed to deal with unpredictable packet loss at actuator nodes. It can significantly improve the QoS of WSANs under unreliable channel conditions, and facilitates the implementation of (mobile) control applications over WSANs.
- The proposed approach is evaluated and verified using trace-based simulations that extract data from the real experiments. In this way, real characteristics of the wireless links are taken into account in performance evaluation. Promising results are presented and analyzed.

The proposed design methodology is a *generic* solution in the sense that: 1) It does not require any modification to low layers such as physical layer, MAC layer, and transport layer within the network protocol stack. Only the application layer is involved. Therefore, it is independent of the underlying communication protocols, for example, MAC and routing protocols, utilized in the WSAN. 2) It does not require any knowledge about the models of the physical systems to be controlled or the design of the control algorithms. It is suitable for a wide range of control applications. 3) The proposed algorithm is computationally cheap yielding only a small runtime overhead. This meets well the general WSAN design requirements stemming from the constraints on data processing capacity and energy consumption in actuator nodes, thus making the proposed approach applicable to various WSAN platforms subject to resource constraints.

This paper is organized as follows. Section 2 briefly reviews related work with respect to WSAN, control over WSNs, link quality analysis, and packet loss handling. Section 3 discusses the architecture and design challenges of WSANs from an application perspective. In Section 4, the properties of link quality are captured in terms of packet loss rate through experiments on a real WSAN. Aiming to improve the reliability and QoS of WSANs, Section 5 proposes a method to handle packet loss. Section 6 evaluates the performance of the proposed approach using trace-based simulations. Finally, Section 7 concludes the paper.

## 2. Related Work

While significant effort has been made in research and development of WSNs in recent years and tremendous advancements have been achieved with respect to deployment, localization, MAC protocols, power control, topology control, routing, distributed signal processing, and security [22], WSANs are a relatively new research area with limited progress. Akyildiz and Kasimoglu [1] described research challenges for coordination and communication problems in WSANs. Rezgui and Eltoweissy [2] discussed the opportunities and challenges for service-oriented sensor/actuator networks. Ngai *et al.* [23] studied the route design problem for mobile actuators and developed a practical algorithm to reduce the waiting time of sensors. Melodia *et al.* [3] presented a sensor-actuator coordination model based on an event-driven partitioning paradigm. Sikka *et al.* [24] deployed a large heterogeneous WSAN on a working farm to explore sensor network applications that can help manage large-scale farming systems. A power-aware many-to-many routing protocol can be found in [9]. Despite their contributions in WSAN, none of the networks designed in these works are particularly for real-time control applications.

Sensor networks have started to attract the attention of control engineers. Kumar *et al.* [10] developed distributed ad-hoc network algorithms to facilitate executing control procedures in a distributed manner. Li [11] prototyped a light monitoring and control application as a case study of WSANs. Oh *et al.* [25] illustrate the main challenges in developing real-time control systems for pursuit-evasion games using a large-scale sensor network. A mixed model for design, analysis, and synthesis of control algorithms within sensor networks has been presented in [26]. Korber *et al.* [27] dealt with some of the design issues of a highly modular and scalable implementation of a WSAN for factory automation applications. Considering networked control systems (NCSs) over WSNs, Nikolakopoulos *et al.* [17] developed a gain scheduler to cope with time-varying delay induced by dynamic changes in the number of hops in multi-hop communications. Witrant *et al.* [18] also considered the effect of time-varying delay caused by multi-hop communication, and proposed a predictive control scheme with a delay estimator. Various design challenges associated with control over wireless networks have been addressed in these papers, but the impact of packet loss as a result of unreliable communications in WSANs, particularly those with mobile nodes, on the performance of the control applications remains an open issue, and needs to be investigated systematically.

Experiments have been conducted for analysis of the link quality in sensor networks, e.g., [28-31]. The authors reported, respectively, their measurements of the packet delivery performance of sensor networks of different sizes in different environments. Some important characteristics with respect to, e.g., packet loss rate in WSANs, have been captured in these papers; but none of the analysis has intended for real-time control applications. For example, the relationship between the resulting control performance and the link quality has not yet been characterised. Also, no methods have been developed in these papers to address the observed unreliable packet delivery.

In the control community, effort has been made for packet loss compensation. A recent survey on this topic can be found in [32]. Despite their differences, most of existing packet loss compensation methods have the common features that: 1) they depend heavily on the knowledge about the accurate models of the physical systems to be controlled, and, possibly, the controller design; and 2) the relevant algorithms are computationally intensive. Due to these reasons, they are impractical for real systems lacking well-established mathematical models. In particular, they are not the desirable solutions for resource-constrained WSANs because of too large computational overheads.

In summary, the field of WSANs is emerging; but the full potential of WSANs for control applications is yet to be explored. For this purpose, the characteristics of the link quality of real-world WSANs in terms of packet loss rate, which may significantly affect the performance of control applications, should be analysed. Resource-efficient paradigms addressing the packet loss problem need to be developed when designing WSANs for (mobile) control applications.

## 3. WSAN for Control Applications

This section describes the architecture of WSANs as a backbone for constructing control applications. The main challenges in design of WSANs will also be discussed briefly.

In general, there are three essential components in a WSAN: sensors, actuators, and base stations. The roles of sensors and actuators have been described previously, while the base stations are often responsible for monitoring and managing the overall network through communications with sensors and actuators. Depending on whether or not there are explicit controller entities within the network, two types of architectures of WSANs for control applications can be distinguished, as shown in Fig. 1 and Fig. 2, respectively. These two architectures are called automated architecture and semi-automated architecture, respectively, in [1].

In the first type of architecture as shown in Fig. 1(a), there is no explicit controller entity in the WSAN. In this case, controllers are embedded into the actuators and control algorithms for making decisions on what actions should be performed upon the physical systems will be executed on the actuator nodes. The data gathered by sensors will be transmitted directly to the corresponding actuators via single-hop or multi-hop communications. The actuators then process all incoming data by executing pre-designed control algorithms and perform appropriate actions. From the control perspective, the actuator nodes serve as not only the actuators but also the controllers in control loops. From a high-level view, wireless communications over WSANs are involved only in transmitting the sensed data from sensors to actuators; control commands do not need to experience any wireless transmission because the controllers and the actuators are logically integrated, as shown in Fig. 1(b).

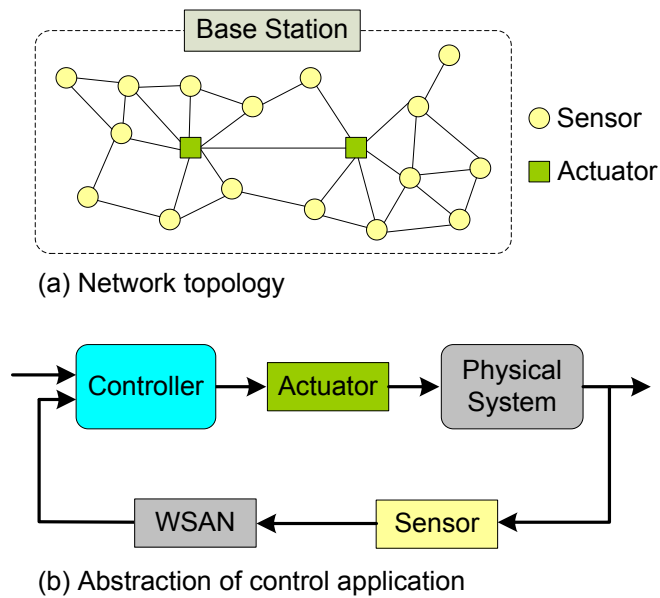

**Figure 1.** WSAN Architecture without explicit controllers.

Fig. 2(a) shows the second type of architecture, in which one or more controller entities explicitly exist in the WSAN. The controller entities could be functional modules embedded in the base stations or separated nodes equipped with sufficient computation and communication capacities. With this architecture, sensors send the collected data to the controller entities. The controller entities then execute certain control algorithms to produce control commands and send them to actuators. Finally, the actuators perform the actions. In this context, both the sensor data and control commands need to be transmitted wirelessly in a single-hop or multi-hop fashion. A high-level view of the applications of this architecture is depicted in Fig. 2(b).

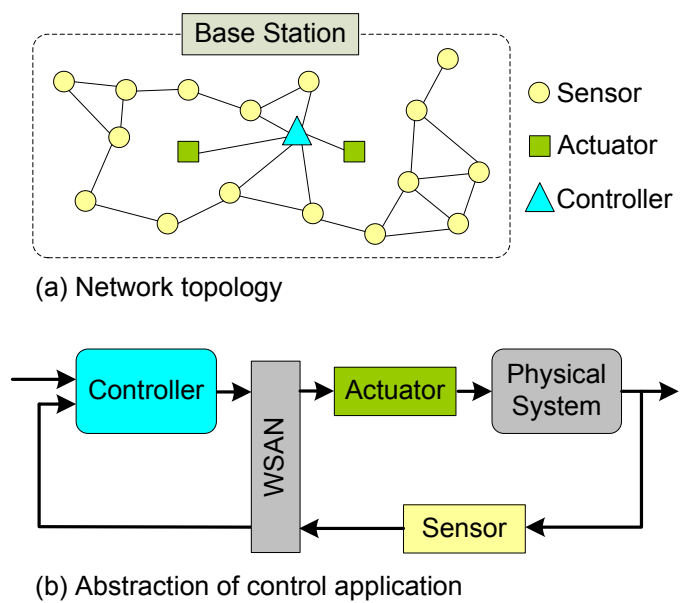

**Figure 2.** WSAN Architecture with explicit controllers.

In combination with the unique characteristics of WSANs, control applications pose the following main challenges associated with the design of WSANs:

- *Reliability*. From the control perspective, packet loss degrades control performance and even causes system instability. Because practical control applications can only tolerate occasional packet losses with a certain upper bound of allowable packet loss rate, WSAN design should minimize the occurrence of packet losses as much as possible. Ideally, every packet should be transmitted successfully from the source to the destination without loss. However, due to many factors such as low-power radio communication, variable transmit power, multi-hop transmission, noise, radio interference, and node mobility, packet loss cannot be completely avoided in WSANs. The challenge then becomes how to improve the reliability of the network system in the presence of packet loss.
- *Real-time constraint*. Control systems are inherently real-time systems in the sense that control actions must be performed on the physical systems by their deadlines. It is worth mentioning that real-time does not necessarily mean 'fast'. For real-time control applications, both delay and its jitter should be limited and predictable in favor of control performance improvement. However, the use of dynamic routing protocols and random MAC protocols (e.g., CSMA/CA), as well as the mobility of nodes, makes the WSAN-induced delay time-varying and unpredictable. The challenge here is how to guarantee the delay is sufficiently small and deterministic with small jitter so that it will not significantly degrade the control performance.

As mentioned in Section 2, WSAN-induced time-varying delay has been addressed in [17,18]. In the recent work by Tian and colleagues [19-21], a real-time queuing protocol has been developed that can be used to tackle the second design challenge of WSANs. Therefore, in this paper, we concentrate our attention on the first design challenge concerning the reliability of WSANs. The existence of mobile nodes in the system undoubtedly makes this task even more difficult.

In the following, we will restrict our description to WSANs with the first type of architecture as shown in Fig. 1, since it is more resource-efficient and is more representative of the next generation of sensor networks. It is, however, noteworthy that our design method is applicable to a wide range of WSANs with arbitrary architectures. Typical examples of (ongoing) real-world application scenarios of the considered WSANs include the pollution source location problem and the fire in a road tunnel scenario where mobile robots must be controlled in a WSAN to accomplish certain jobs [33]. When illustrating our approach, we will exploit a high-level abstraction of the WSANs for various application setups in order to maintain the application independency and wide applicability of our solution.

## 4. Experimental Analysis of Link Quality

In order to address the challenge of unreliable communication in WSANs, it is necessary to understand first how unreliable practical WSANs really are. That is, the packet loss behavior of the network should be studied. This is done in this work through conducting experiments on a real WSAN system and collecting quantitative data to capture the channel characteristics. This section reports our experimental measurements that characterize the link quality in terms of packet loss rate of a practical WSAN, since in this context the packet loss rate is a critical factor affecting the performance of the control applications. Instead of an exhaustive study of link quality with respect to a lot of factors, e.g., platform, environment, deployment, time, etc., which has been done e.g. in [27-30], we focus on a simple yet sufficiently illustrative characterization of the link quality that help make decisions concerning WSAN design for mobile control applications.

*4.1. Experimental Setup*

The sensor nodes used in the experiments is the MICA2 motes from Crossbow [34]. The MICA2 mote, designed specifically for deeply embedded sensor networks, is based on the Atmel ATmega128L microcontroller. Each sensor node contains 128KB program flash memory, 512KB measurement flash, and 4KB configuration EEPROM. MICA2 uses the Chipcon CC1000 wireless transceiver, and supports multiple channels (868/916 MHz), hardware encoding (Manchester), frequency shit keying (FSK) modulation, and up to 38.4 kbps data rate. The RF power of MICA2 is programmable from -20 to +5 dBm, and the receive sensitivty is -98 dBm. The MICA2 51-pin expansion connector supports analog inputs, digital I/O, I2C, SPI and UART interfaces. These interfaces make it easy to connect to a wide variety of external peripherals. Any MICA2 Mote can function as a base station when it is connected to a standard PC (personal computer) interface or gateway board. A base station allows the aggregation of sensor network data onto a PC or other computer platform.

MICA2 runs TinyOS, an open-source embedded operating system developed at UC Berkeley. It provides basic system services, such as communication and simple process scheduling, and access to hardware components such as sensors and actuators. The MAC layer implements a simple CSMA/CA protocol. A link-level acknowledgement can be sent by the receiver for each successful packet.

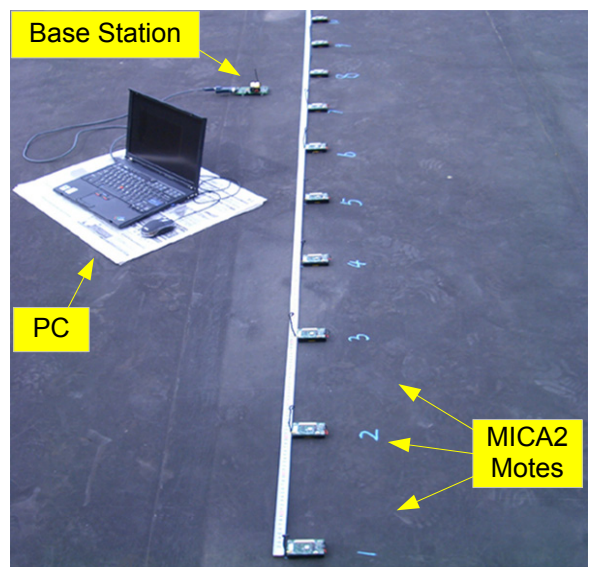

**Figure 3.** Experimental deployment.

The experiments are conducted on an open ground, as shown in Fig. 3. Nodes are placed equidistantly along a line with a spacing of 0.5m. One PC is connected with the base station using a MIB510 mote interface board. The PC is used to configure network parameters and collect experimental data via the base station. In the experiment, the first node, i.e. the mote on the position marked as '1' in Fig. 3, transmits continuously a 13 byte data packet at a rate of 8 packets per second.

The packet loss rates associated with the remaining nodes with different distances from the first node are measured.

*4.2. Observations*

In a WSAN with mobile (actuator) nodes, the distance between the mobile actuator and the sensor that are involved in transmitting the sensed data will change over time, thus aggravating the variability of link quality. This is why we pay a special attention to examining how the link quality varies over distance.

Fig. 4 shows the packet loss rates on different nodes when the transmit power is set to 0dBm. Within the distance of 7m, the packet loss rates remain less than 10%. Beyond 30m, the packet loss rates are fairly close to 100%, implying that almost all packets sent to the nodes are lost. This indicates that the radio range is approximately 30m. For links with distances ranging from 7 to 30m, the packet loss rates could vary drastically. For example, the packet loss rates vary nearly from 0% to 100% in the area between 9 and 13m. It can be observed that nodes far away from the transmitter may possibly undergo smaller packet loss rates than those near the transmitter.

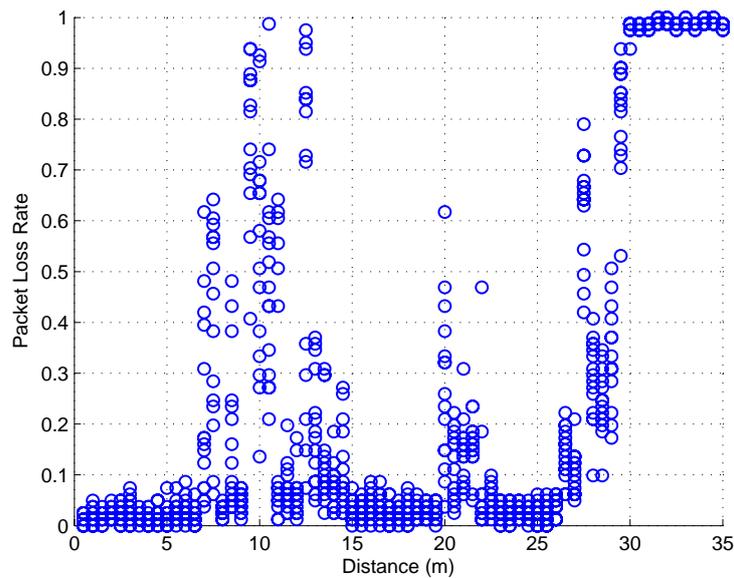

**Figure 4.** Packet loss rate versus distance (0dBm).

Fig. 5 plots the recorded data when the transmit power is set to be -5dBm. Similarly, the packet loss rates show great variability and irregularity both over different distances and at a given distance. Compared to the higher power case shown in Fig. 4, almost 100% packet loss rates are observed at a smaller distance, and the radio range decreases to 26m.

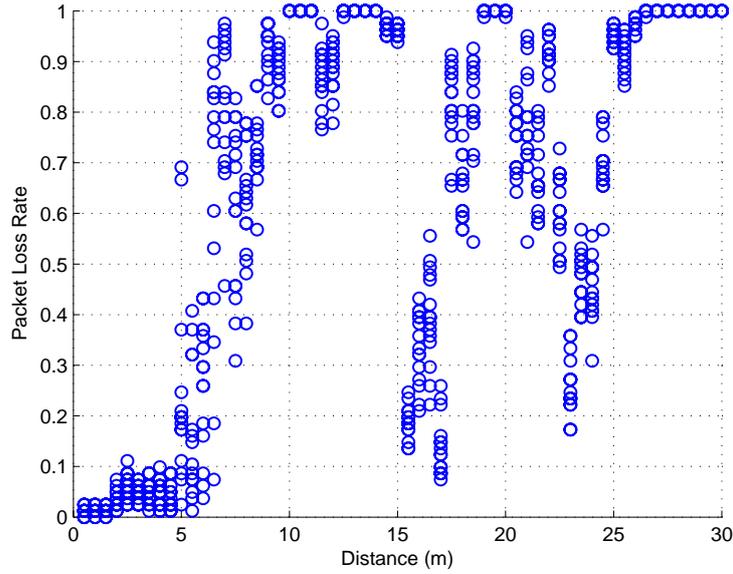

**Figure 5.** Packet loss rate vs. distance (-5dBm).

We have observed that the packet loss rates over the wireless channel in real WSANs are highly variable and irregular. The assumptions that link quality is exclusively based on distance, which is often made when modelling the link quality of sensor networks [35,36], simplify system analysis but can be practically questionable. This could be further justified by the fact that the link quality of WSANs depends also on many factors other than distance and transmit power, such as the environment, noise, radio frequency, modulation scheme, and (hardware) platform in use, just to mention a few. Due to the unpredictability of the link quality in terms of packet loss rate and the impracticability to model it accurately, it is now imperative to develop a platform-independent paradigm to enhance the reliability of WSANs under lossy conditions. A desirable solution should be widely applicable to diverse application scenarios with different system and environment setups.

## 5. Dealing with Packet Loss on Actuators

To meet the above requirement, we attempt to develop an application-level design methodology for WSANs in mobile control applications. The principles in our development are: 1) to modify only the application layer of the networks without exploiting any application-specific (lower layer) network protocols, 2) not to use any statistic information about the distribution of packet loss rate in any specific WSAN, and 3) not to use the knowledge about the models of the controlled physical systems and the controller design of the target control applications.

We propose to use a simple yet efficient method on the actuator nodes to cope with packet loss occurring in WSANs. The basic idea is: whenever a sensor data packet is lost, the actuator will still produce a control command (usually called *control input* in control terms) by means of prediction from previous control command values.

For a control loop within a (possibly large-scale) system shown in Fig.1, suppose that the $k$-th sensor data is lost. In this case the actuator (to which the last data should be sent) will calculate an estimate of the control command using the PID (proportional-integral-derivative) algorithm, the most popular control algorithm in the control community, as follows:

$$\hat{u}(k) = \tilde{K}_P u(k-1) + \tilde{K}_I \sum_{i=k-m}^{k-1} \frac{u(i)}{m} + \tilde{K}_D (u(k-1) - u(k-2)) \tag{1}$$

where $\hat{u}(k)$ is the estimate of the $k$-th control command $u(k)$, $\tilde{K}_P$, $\tilde{K}_I$, $\tilde{K}_D$ and $m$ are user-specified parameters. Using (1), the actuator predicts $u(k)$ based on the previous $m$ consecutive control commands (which are also possibly predicted values) in the case of packet loss and performs the actions corresponding to the value of $\hat{u}(k)$. Given that the accuracy of the prediction of control commands is sufficiently high, proper actions will be performed on the controlled physical system in every sampling period, regardless of the loss of the sensor data. In this way, the effect of packet loss on the performance of the control applications can be substantially reduced. In other words, the reliability of the WSAN is improved, from the application point of view.

The PID algorithm is here used to predict the unavailable control commands that result from sensor packet loss. In the control community, in contrast, it is typically used in control system design, as will be shown later in Section 6. Varma *et al.* [37] have used a similar method called nqPID to predict CPU workload in dynamic voltage scaling systems. It proved quite effective and insensitive to parameter changes.

The work flow of the actuator can be illustrated as follows:

**Input**: Sensor data  
**Output**: Control command  
*Begin*  
    **If** *the sensor data is lost* **then**  
        Compute $\hat{u}(k)$ using (1)  
    **Else**  
        Compute $u(k)$ using pre-designed control algorithm(s)  
    **End if**  
    Store $u(k)$ or $\hat{u}(k)$ in memory  
    Discard $u(k-m)$ in the memory  
    Perform actions corresponding to $u(k)$ or $\hat{u}(k)$  
*End*

It can be seen that this design method is quite simple. The major overhead is a small fraction of memory to temporarily store the previous $m$ control commands. Despite this, it does not depend on any knowledge about the underlying platform, environment, link quality characteristics, models of the controlled systems, or controller design. Furthermore, only a very limited amount of computations have been introduced, which fulfills well the general requirements of WSANs concerning the constraints on computational capacity and energy expenditure. In addition, it is worth mentioning that although zero delay is assumed in this paper, the proposed design method can be easily combined with the real-time queuing protocol developed in [19-21] to deal with simultaneously time-varying delay and packet loss.

## 6. Performance Evaluation

In this section, we conduct trace-based simulations using Matlab to evaluate the performance of the above-proposed design methodology for WSANs.

*6.1. Control Application Overview*

In the simulations, a commonplace control system design is used to keep the results as general as possible. The model of the controlled physical system is given below, which may represent an inverted pendulum system, a common benchmark problem in the control field [38]:

$$G(s) = \frac{4.546s}{s^3 + 0.182s^2 - 31.182s - 4.454}$$

The controllers use the PID control law, the most popular control law in practical control applications, with the following parameters: $K_P = 120$, $K_I = 1000$, and $K_D = 5$. The PID control algorithm is implemented in the actuator as follows [14]:

$$e(k) = r(k) - y(k)$$
$$P(k) = K_P e(k)$$
$$I(k) = I(k-1) + K_I h(e(k) + e(k-1))/2$$
$$D(k) = K_D (e(k) - e(k-1))/h$$
$$u(k) = P(k) + I(k) + D(k)$$

where $r(k)$ is the desired system output (i.e., reference input or set-point), $y(k)$ is the sensed value of the system output (i.e., measurement), $h$ is the sampling period of the sensor. In the simulations, $h$ is set to 20ms.

As the major perturbations on the controlled system, the reference input changes over time as a square wave with a period of 2s. To measure the performance of the control application, the integral of absolute error (IAE), one of the widely used control performance metric defined as $J(t) = \int_0^t |r(\tau) - y(\tau)| d\tau$, is recorded. The bigger the IAE value the worse the control performance.

*6.2. Simulation Results and Analysis*

Because of the mobility of the actuator node, the distance between the sensor and the actuator varies during runtime according to Fig. 6. The RF power of the sensor is assumed to be 0dBm. Accordingly, the packet loss rates with respect to different distances will be randomly extracted from the data set reported in Fig. 4. For instance, when the distance is 10m from $t = 16$ to 20s, the packet loss rates will take random values out of {0.7160, 0.2716, 0.6790, 0.9136, 0.9259, 0.6543, 0.3827, 0.6543, 0.4691, 0.3333, 0.2963, 0.1358, 0.5062, 0.6790, 0.5802}.

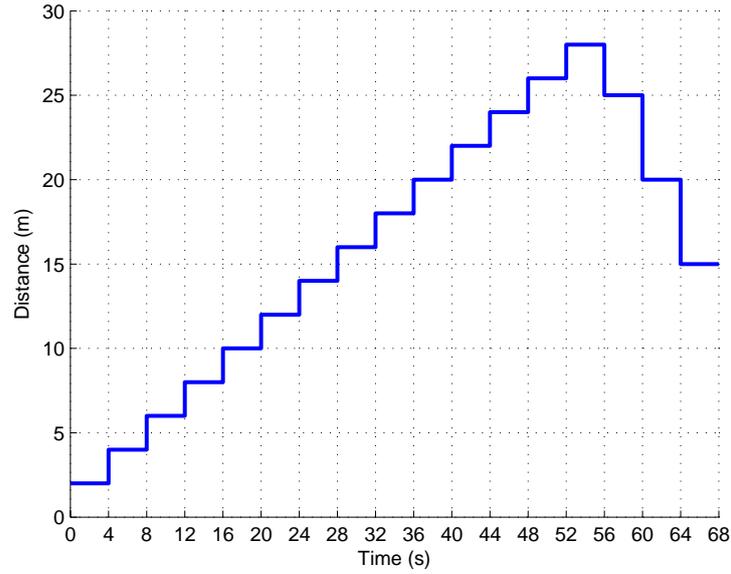

**Figure 6.** Variable distance in simulations.

The performance of the control application is compared with respect to two different WSAN design methods: 1) traditional design method without packet loss handling mechanism on the event-triggered actuator; and 2) the application-level design methodology proposed in this paper. Some relevant parameters are set as follows: $\tilde{K}_P = 0.3$, $\tilde{K}_I = 0.2$, $\tilde{K}_D = 0.5$, and $m = 3$.

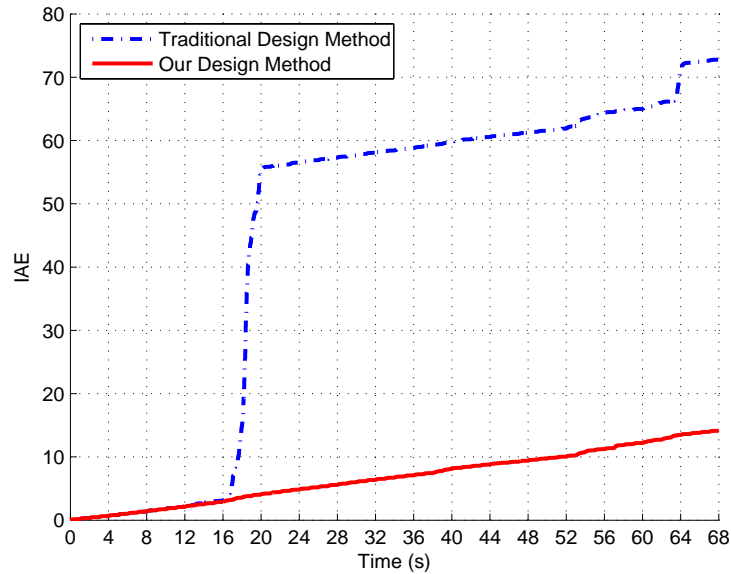

**Figure 7.** Control performance in terms of IAE.

Fig. 7 shows the control performance in terms of the IAE associated with different design methods. The cumulated IAE value of the system designed using our method is only 19.4% that of the system using traditional design method. The rapid increase in IAE from time $t = 16$s to 20s implies that the system becomes unstable during this period of time. This can also be seen from Fig. 8 (the upper part), where the measured/sensed system output is depicted. The instability is mainly caused by the large packet loss rates at the distance of 10m, which has been shown in Fig. 4.

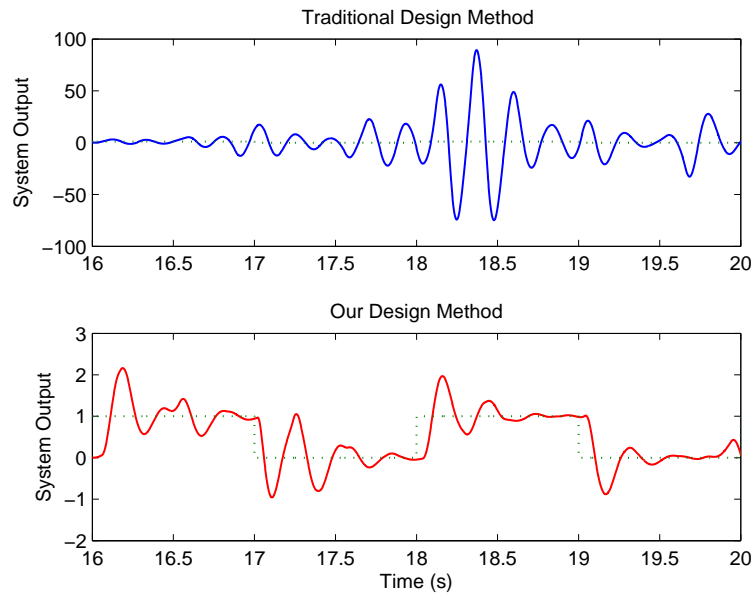

**Figure 8.** System output.

In contrast, the system remains stable all the time, when the design method proposed in this paper is employed. This is justified by the quite slow increase in IAE throughout the simulation, see Fig. 7. As also shown in Fig. 8 (the lower part), the performance of the control application is satisfactory even when the system may encounter considerably severe packet loss. Since no nodes are allowed to work beyond its radio range (i.e., 30m in this case) in practice, the above results demonstrate that the method proposed in this work is effective in mobile control applications where the communication distance may change over time.

## 7. Conclusion

This paper deals with the design of WSANs for control applications. The related design challenges have been discussed with respect to reliability and real-time constraints. With focus on improving the reliability of WSANs to provide control applications with network QoS guarantees, a generic application-level design methodology has been presented. The link quality of WSANs has been examined in terms of packet loss rate through experimenting on a real WSAN system. From the experimental observations, a simple yet effective method has been developed to deal with unpredictable packet loss on the actuator nodes. The proposed design methodology has also been verified through trace-based simulations. It enables mobile control applications over WSANs since it can guarantee satisfactory control performance even in the presence of significant packet loss.

The design methodology proposed in this paper is independent of the computation and communication platforms upon which the WSAN is built, and the environment in which the WSAN is deployed. Also, it does not rely on the system models and controller design of the target control applications. Furthermore, it is computationally cheap since only a small overhead is introduced.

Therefore, the proposed design methodology can be applied in a wide range of WSAN-based control applications.

**Acknowledgements**

Authors Xia and Tian would like to thank Australian Research Council (ARC) for its support under the Discovery Projects Grant Scheme (grant ID: DP0559111). The authors are grateful to Jing Yu at Zhejiang University for her assistance in collecting experimental data.